\documentclass[twocolumn,english,prl]{revtex4-1}
\usepackage[T1]{fontenc}
\usepackage[latin9]{inputenc}
\usepackage{color}
\usepackage{babel}
\usepackage{graphicx}
\usepackage{esint}
\usepackage[unicode=true,pdfusetitle,
 bookmarks=true,bookmarksnumbered=false,bookmarksopen=false,
 breaklinks=false,pdfborder={0 0 0},pdfborderstyle={},backref=false,colorlinks=true]
 {hyperref}

\makeatletter

\usepackage{epsfig}\usepackage{color}\usepackage{booktabs}
\usepackage{hyperref}
\hypersetup{colorlinks=true,citecolor=blue}

\makeatother

\begin{document}

\title{Aging is a (log-)Poisson Process, not a Renewal Process }

\author{Stefan Boettcher$^{1}$, Dominic M. Robe$^{1}$, and Paolo Sibani$^{2}$}

\affiliation{$^{1}$Department of Physics, Emory University, Atlanta, GA 30322,
USA~\\
$^{2}$Institut for Fysik Kemi og Farmaci, Syddansk Universitet, DK5230
Odense M, Denmark}
\begin{abstract}
Aging is a ubiquitous relaxation dynamic in disordered materials.
It ensues after a rapid quench from an equilibrium ``fluid'' state
into a non-equilibrium, history-dependent jammed state. We propose
a physically motivated description that contrasts sharply with a continuous-time
random walk (CTRW) with broadly distributed trapping times commonly
used to fit aging data. A renewal process like CTRW proves irreconcilable
with the log-Poisson statistic exhibited, for example, by jammed colloids
as well as by disordered magnets. A log-Poisson process is characteristic
of the intermittent and decelerating dynamics of jammed matter usually
activated by record-breaking fluctuations (``quakes''). We show
that such a record dynamics (RD) provides a universal model for aging,
physically grounded in generic features of free-energy landscapes
of disordered systems.
\end{abstract}
\maketitle
\begin{figure*}
\vspace{-0.0cm}

\hfill{}\includegraphics[viewport=0bp 400bp 780bp 600bp,clip,width=1\textwidth]{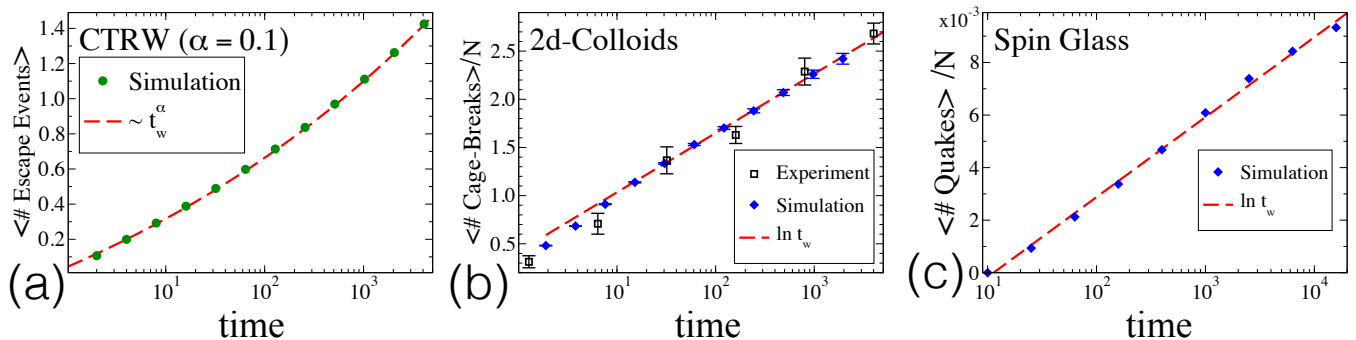}
\hfill{}

\vspace{-0.5cm}

\caption{\label{fig:Rate}Accumulated number of relaxation events (``quakes'')
with time $t$ after a quench during simulations of (a) a continuous-time
random walk (CTRW), (b) a jammed \emph{2d} colloids, and (c) a \emph{3d}
Edwards-Anderson spin glass. In (a), we sampled the number of escapes
by evolving Eq.~(\ref{eq:CTRWpsi}) for a rather small value of $\alpha=0.1$,
to more closely resemble the physical aging data in (b-c). In (b)
we measured intermittent cage-breaks after a quench, marked by irreversible
neighborhood swaps in the molecular-dynamics simulations. Also shown
are the data from the colloidal experiment by Yunker et al~\cite{Yunker09,Robe16}.
In (c), we obtained the irreversible barrier crossings in a \emph{3d}
Edwards-Anderson spin glass. Note that the physical data is growing
logarithmically with time, while such a behavior is obtained in the
CTRW only for $\alpha\to0$.}
\end{figure*}

During a quench, when temperature is dropped in a complex fluid or
disordered magnet~\cite{Lundgren83,Jonsson1995,Rodriguez03,Vincent06,Parker03,Rodriguez13},
density is rammed up in a colloidal system~\cite{Hodge1995,Weeks00,Courtland03,Yunker09},
or strain is intensified on a granular pile~\cite{Lahini2017,Nowak98},
amorphous materials begin to jam up such that relaxational timescales
exceed experimental capabilities.In the ensuing aging process~\cite{Struik78},
observables retain a memory of the time $t$ since the quench at $t=0$,
signifying the breaking of time-translational invariance and the non-equilibrium
nature of the state. Similar phenomena have been observed also in
protein dynamics~\cite{Hu2015}, friction \cite{Dillavou18}, or
financial time series~\cite{Cherstvy2017}. In a jammed disordered
system, structural relaxation towards some far-removed equilibrium
state proceeds by increasingly rare, activated events~\cite{Bissig03,Sibani06a,Yunker09,Kajiya13,Tanaka17},
see Fig.~\ref{fig:Rate}. Furthermore, two-time measures of macroscopic observables, 
taken for some time $\Delta t=t-t_{w}$ starting
at $t_{w}$, scale most reasonably as a function of the single variable $t/t_{w}$.
(In contrast, an equilibrium process would remain time-translational
invariant, depending on $\Delta t$ alone.) For example, experimental
data for the thermo-remanent magnetization of a glassy magnet~\cite{Rodriguez03,Sibani06a},
or for the persistence and the mean-square displacement of colloidal
particles~\cite{Robe16} as well as their entire displacement probability
density (van Hove) function\ \cite{Robe18}, collapse when plotted
as function of $t/t_{w}$. The wide variety of metastable systems exhibiting
this phenomenology suggests its universality~\cite{Cipelletti00,Chaudhuri07}
and call for a unified coarse-grained description, independent of microscopic
detail~\cite{Becker14,Amir2012,Schulz2014}. 

Observations like this have become the basis for models that treat
aging as a renewal process~\cite{Bouchaud92,Burov2010,Schulz2014,Metzler2014,Metzler2015}.
In the trap model~\cite{Bouchaud92}, for instance, the entire system
performs a random walk through a configuration space filled with traps
possessing a broad (power-law) distribution of escape times. Thus,
the older the system, encountering ever ``deeper'' traps will become
more likely, and the deepest trap encountered dominates all previous
timescales. However, once escaped, no memory of previous events informs
future events. As noted in Ref.~\cite{Sibani13}, applying this type
of description to aging systems violates the system size scaling and
self-averaging properties of macroscopic variables, which are universally
observed in nature. The problem is avoided in continuous-time random
walks models (CTRW), where each particle in a colloid, say, is now
endowed with a power-law distribution of times between displacements.
Over time, particles perform intermittent jumps, interpreted as them
breaking out of their cages formed by surrounding particles, as particle-tracking
observations tend to justify~\cite{Weeks00,Chaudhuri07}. In this
Communinication we show, however, that any such renewal process is
ruled out as an underlying  physical mechanism by demonstrating that 
the transitions from one metastable state to the next  
follow a log-Poisson process which originates from the record-sized
fluctuations needed to relax the aging system. 

\begin{figure*}
\vspace{-0cm}

\hfill{}\includegraphics[viewport=0bp 400bp 792bp 600bp,clip,width=1\textwidth]{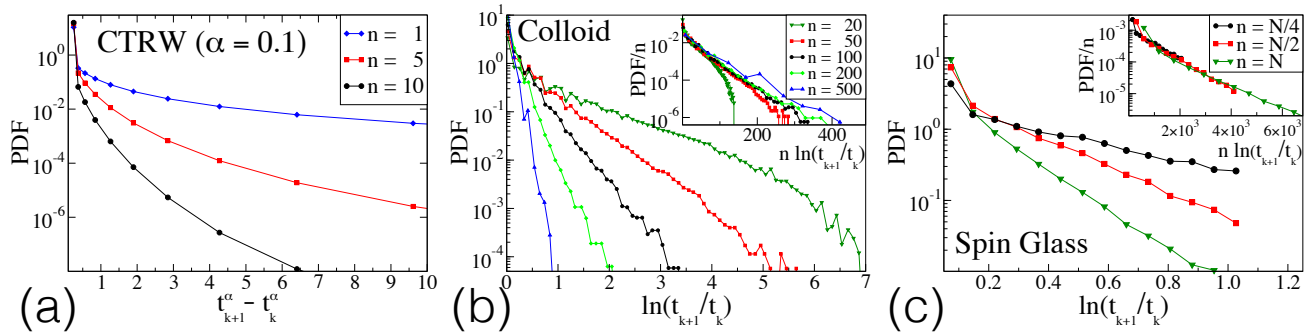}\hfill{}

\vspace{-0.2cm}

\caption{\label{fig:logPoisson} Poisson statistics of (a) the CTRW in Eqs.~(\ref{eq:CTRW}-\ref{eq:CTRWpsi})
for $\alpha=0.1$, and for the aging dynamics in simulations of (b)
a colloid and (c) a spin glass. Times $t_{k}$ mark the $k^{{\rm th}}$
event, i.e., escaping a trap in the CTRW, an irreversible cage-break
in a colloid, or quakes in a spin glass. The CTRW does not have the
exponential form expected for a Poisson process, while both physical
aging processes do. Generally, when too many events (i.e., too many
degrees of freedom) are blurred together, it hides the local impact
of the decelerating activated events and tails weaken. Record dynamics
predicts that the rate of events is proportional to the number $n$
of degrees of freedom observed. As the inset for both, (b) and (c),
shows, the data in the aging simulations indeed collapse when rescaled
by $n$.}
\end{figure*}

By their very nature, renewal processes seem antithetical to aging
because each renewal resets the history. However, it is the broad
distribution of escape times that ensnares an increasing number of
particles for indefinitely long times after the quench and overall
activity decelerates. To be specific, the probability for observing
$m$ such cage-breaking events after the waiting-time $t_{w}$ since
the quench in a single CTRW is given by
\begin{equation}
p_{m}\left(t_{w}\right)=\int_{0}^{t_{w}}d\tau\,p_{m-1}\left(t_{w}-\tau\right)\,\psi\left(\tau\right),\label{eq:CTRW}
\end{equation}
initiated with $p_{m}\left(0\right)=\delta_{m,0}$, and a inter-event-time
distribution
\begin{equation}
\psi\left(\tau\right)\sim\tau^{-1-\alpha},\qquad\left(0<\alpha\leq1\right).\label{eq:CTRWpsi}
\end{equation}
For such $\alpha$, inter-event times, i.e., escape times, have a
diverging mean and the rate at which events are observed, $\partial_{t}\left\langle m\right\rangle \sim t_{w}^{-1+\alpha}$,
indeed decelerates, so that the accumulated number of events rises
sub-linearly, see Fig.~\ref{fig:Rate}a. Moreover, when such a process
has evolved up to time $t_{w}$ after the quench, the probability
to observe the next (escape) event within a time-interval $\Delta t=t-t_{w}$
exhibits the $t/t_{w}$-dependence~\footnote{
At the level of macroscopic observables, this $t/t_{w}$-dependence is however only approximate. Macroscopic data are often empirically scaled using
`sub-aging', which is a cross-over phenomenon~\cite{Warren13} on top of the Record Dynamics mechanism, as discussed in ``Origin of `end of aging' and sub-aging scaling behavior in glassy dynamics''~\cite{Sibani10}.}
characteristic of most aging phenomena~\cite{Schulz2014}:
As the time needed to escape the typical trap entered at $t_{w}$
is itself $\propto t_{w}$, then so is $\Delta t\propto t_{w}$, which
constitutes some fraction of the total escape time. These are powerful
features of renewal models that have contributed greatly to justify
their widespread application to fit data produced in a wide variety
of aging experiments~\cite{Chaudhuri07,Cherstvy2017,Metzler2014,Hu2015,Vincent95,Bouchaud92,Brokmann2003}.
The exponent $\alpha$ provides a readily available parameter to fit
data. However, the \emph{physical} origin of $\alpha$ or $\psi(\tau)$
remains obscure. 

\begin{figure}
\hfill{}\includegraphics[viewport=0bp 130bp 640bp 600bp,clip,width=1\columnwidth]{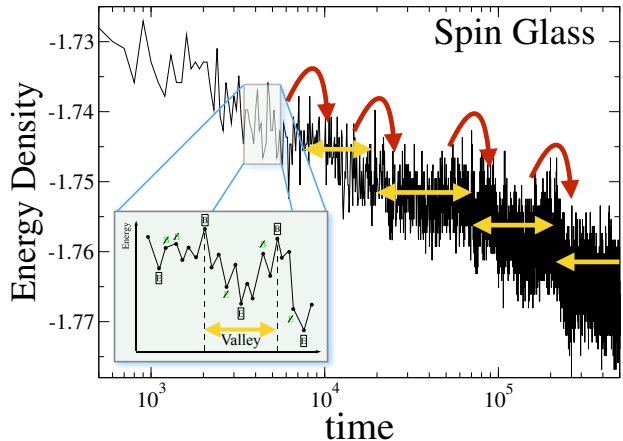}\hfill{}

\vspace{-0.4cm}

\caption{\label{fig:Landscape}Energy trace during a single aging process in
a \emph{3d} EA spin glass system with time $t$ after a quench. The
energy decreases logarithmically with time in a widely fluctuating
manner. Unlike in a renewal process, a glass never returns to energy
levels visited decades earlier, signifying the gradual but significant
structural evolution in the configuration of spins. Energy leaves
the systems in intermittent and \emph{irreversible} escape events ("quakes") 
out of ``valleys'' (yellow arrows) triggered by a record fluctuation
(red arrows). Inset: Tumbling through a complex energy landscape,
a time sequence of lowest energy (E) and highest barrier (B) records
(relative to the most recent \textquotedblleft E\textquotedblright )
is produced~\cite{Dall03,BoSi}. Only highest and lowest records
of the \textquotedblleft E and \textquotedblleft B\textquotedblright{}
are kept to give a strictly alternating sequence \textquotedblleft EBEBE...\textquotedblright .
Then, any \textquotedblleft BEB'' sequence demarcates entering and
escaping a valley. Like in-cage rattle within a colloid, each valley
represents a local meta-stable domain in the landscape where the system
exhibits quasi-equilibrium behavior on time-scales shorter than the
escape time.}
\end{figure}

In light of its potential benefits, it is most instructive to compare
CTRW with a direct measure of the sequence of quake events in realistic
aging processes, to assess their statistical properties more fully.
This we have undertaken in molecular-dynamics simulations of a \emph{2d}
system of bi-disperse colloidal particles and a Monte Carlo simulation
of $3d$ Edwards-Anderson spin glasses. (Details of these simulations
are described in Refs.~\cite{Robe18,Sibani18}). Each provides a
canonical model of aging (and glassy behavior generally) for quite
distinct disordered materials~\cite{Hunter12,F+H}. In a colloid,
disorder is merely structural, arising from the irregular random packing
of the particles~\cite{Hunter12}. In contrast, in a spin glass,
dipolar magnets are localized in a lattice of a-priori fixed but randomly
chosen couplings with their neighbors that would frustrate their optimal
alignment even in any conceivable -- yet dynamically inaccessible
-- equilibrium arrangement~\cite{F+H}. In both systems we follow
the sequence of events after a quench, achieved by either rapidly
expanding the colloidal particles to transition from a low-density
liquid into a high-density jammed state, or by lowering temperature
of the spin glass well below its glass-transition temperature. Sampling
over repeated simulations, we have recorded either the cage-breakings
or the energy barrier-crossings (see Fig.~\ref{fig:Landscape}),
which constitute the irreversible events (or jumps) signifying activated
relaxation in the respective systems. 

In all three cases the rate of events does decelerate as a power-law
with time. However, in both simulations the rate is essentially hyperbolic,
$\partial_{t}\left\langle m\right\rangle \sim t_{w}^{-1}$, such that
the accumulated number of events increases logarithmically with time,
as plotted in Fig.~\ref{fig:Rate}. This feature of the physical
data a renewal process only achieves in the limit of $\alpha\to0$.
That limit is somewhat singular because the distribution in Eq.~(\ref{eq:CTRWpsi})
would become un-normalizable. Arguably, this could be accounted for
within numerical accuracy of the data by stipulating logarithmic factors
or by simply assuming some small value of $\alpha$ (like $\alpha=0.1$,
as used in Fig.~\ref{fig:Rate}a). Yet, the following will demonstrate
that the key discrepancy between a renewal model and the data arises
from the actual \emph{sequence} of quake events. 

Focussing first on the numerical data, we let $t_{k}$ denote the
$k^{{\rm th}}$ quake in the time series of measured irreversible
events extracted from a trajectory. Figs.~\ref{fig:logPoisson}b-c
show the statistics of the logarithmic time differences $\Delta\ln=\ln t_{k+1}-\ln t_{k}=\ln\left(t_{k+1}/t_{k}\right)$,
which we treat as identically distributed stochastic variables.
To wit, to a good approximation, the statistics is described by an exponential
probability density function (PDF) $P_{\Delta\ln}\left(x\right)=\exp\left\{ -x/\mu_{q}\right\} /\mu_{q}$,
which is shown for a number of different system sizes. The rate of
events, $1/\mu_{q}$, increases with system size or simply the number
of tracked particles $n$. That the scaling is indeed linear is shown
in the insets, were all data are collapsed by the scaling transformation
$\Delta\ln\to\Delta\ln/n$. Thus, the form of the data supports the
hypothesis that quake events have a log-Poisson statistic, i.e., the
data follow a Poisson distribution whose only parameter -- the average
-- grows \emph{logarithmically} with time, i.e., $\mu_{q}\propto\ln t$. 

The additivity of Poissonian variables now ensures that the number
of events is an extensive variable. In CTRW, the number of on-going
independent random walks is a fitting parameter rather than a dynamical
consequence of the size of the system at hand. One can nevertheless
look at the sequence $t_{k}$ of renewal events generated by $n$
different random walks. For $n=1$, the PDF of the time intervals
$\tau=t_{k+1}-t_{k}$ between consecutive events at times $t_{k}$
and $t_{k+1}$ by definition reproduces the power-law distribution
$\psi\left(\tau\right)$ in Eq.~(\ref{eq:CTRWpsi}). For $n>1$,
the data from the renewal process retain a power-law distribution,
inconsistent with a Poisson process on any timescale. Even if we define
a generalized $\alpha$-Poisson process, based on the observation
that the rate of events observed in a renewal process is $\partial_{t}\left\langle m\right\rangle \sim t^{-1+\alpha}$,
i.e., $\left\langle m\right\rangle (t,t_{w})\sim t^{\alpha}-t_{w}^{\alpha}$,
the $\alpha$-intervals between events, $t_{k+1}^{\alpha}-t_{k}^{\alpha}$,
remain power-law distributed and distinctly non-Poissonian, as demonstrated
in Fig.~\ref{fig:logPoisson}a. Thus, a renewal process must be rejected
as a model for aging! With further extensions \cite{Lomholt2013},
a non-renewal CTRW can be designed that evolves many walkers in parallel,
one for each future event. Here the system, after leaving the trap of one
walk, immediately enters the already extant trap of another walk,
instead of undergoing renewal. While this model provides a log-Poisson
statistic for all $0<\alpha<1$, any connection to the physical processes
we discuss here is tenuous, at best.

In contrast, it is quite fruitful to view aging as a record dynamics
(RD)~\cite{Sibani93a,Anderson04,SJ13}. In a statistic of records,
a sequence of $t$ independent random numbers drawn from \emph{any}
smooth probability density function produces a record-sized number
at a rate $\partial_{t}\left\langle m\right\rangle \sim1/t$. Hence,
the number of random events tallied between a time $t_{w}$ and $t=t_{w}+\Delta t$
is $\left\langle m\right\rangle (t,t_{w})\sim\ln(t)-\ln(t_{w})\sim f\left(t/t_{w}\right)$.
Thus, RD leads to a log-Poisson statistic~\cite{SJ13}, as found
for the physical data above. RD also considers anomalously large events,
like cage breaks in colloids, as being essential to substantially
relax the system and having to be viewed as distinct from the Gaussian
fluctuations of in-cage rattle. (This distinction is also essential
to CTRW, as the discussion before Eq.~2 in Ref.~\cite{Chaudhuri07}
shows.) Yet, such a relaxation must entail a structural change --
the physical essence of aging -- that makes subsequent relaxation
even harder, see Fig.~\ref{fig:Landscape}. For example, to facilitate
a cage-break, a certain number of surrounding particles have to ``conspire''
to move via some rare, random fluctuation~\cite{Yunker09}. For that
event to qualify as an irreversible loss of free energy, the resulting
structure must have increased stability, however marginal. A subsequent
cage-break therefore requires even more particles to conspire. With
each fluctuation being exponentially unlikely in the number of particles,
cage-breaks represent \emph{records} in an independent sequence of
random events that ``set the clock'' for the activated dynamics,
resulting in the observed log-Poisson statistics. Then, any two-time
observable becomes \emph{subordinate}~\cite{Sibani06a} to this clock:
$C\left(t,t_{w}\right)=C\left[\left\langle m\right\rangle (t,t_{w})\right]=C\left(t/t_{w}\right)$.
Indeed, much of the experimental colloidal tracking data in the aging
regime can be collapsed in this manner\ \cite{BoSi09,Robe16}. In
our colloidal simulations this is verified by the mean-square displacement
(MSD) of particles between times $t_{w}$ and $t=t_{w}+\Delta t$
after a quench at $t_{w}=0$ which similarly collapses onto a single
function of $t/t_{w}$, as shown in Ref.\ \cite{Robe18} and previously
obtained for experiments in Ref.~\cite{Robe16}. That function is
consistent with a logarithmic growth of MSD with $t/t_{w}$, indicative
of jumps caused by activations in a decelerating sequence of record
events. In fact, the entire van-Hove function for particle displacements
can be collapsed in this manner~\cite{Robe18}. RD furthermore predicts
that the rate of events is proportional to the number $n$ of particles
observed, as verified by the data. Thus, RD provides an effective
model of aging, similarly devoid of microscopic details and hence
apt to capture the universality of aging. 

Beyond being an effective model, RD allows deep physical insights
into the aging dynamics. While ordinary exponential relaxation in
the total energy implicates a gradual descend of the material through
a smooth (convex) landscape, the temporal and spatial heterogeneity
(i.e., intermittency and ``dynamic heterogeneity'') observed during
aging is indicative of a non-convex landscape with many local minima
trapping the system in a hierarchy of meta-stable states. However,
unlike in a renewal process, leaving such a trap is an irreversible
process that has memorable consequences. Free energy (e.g., free volume
in the case of a hard-sphere colloid) leaves the system which gets
deeper entrenched, calling for a progression of record fluctuations,
see Fig.~\ref{fig:Landscape}. 

\begin{figure}
\hfill{}\includegraphics[viewport=80bp 20bp 610bp 720bp,clip,angle=270,width=0.9\columnwidth]{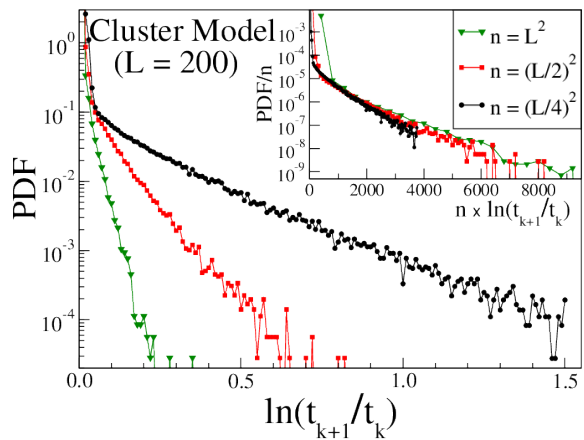}\hfill{}

\vspace{-0.2cm}

\caption{\label{fig:ClusterPoisson}Log-Poisson statistics for the cluster
model\ \cite{Becker14} on a square lattice of size $L=200$. Inset:
Data collapse when rescaled by $n$, as in Fig.\ \ref{fig:logPoisson}b-c
for colloid and spin glass.}
\end{figure}

A phenomenological cluster model of aging has been proposed recently,
based on record dynamics~\cite{Becker14}. It is an on-lattice model
that captures the combined temporal \emph{and} spatial heterogeneity
found in a colloidal system: mobile particles accrete into jammed
clusters only to be re-mobilized in a chance fluctuation (i.e., a
quake in free volume) after a time \emph{exponential} in the size
$h$ of any cluster which occurs with probability $P(h)\sim e^{-h}$.
Following a quench at $t=0$, clusters form and break up to \emph{irreversibly}
distribute their particles to neighboring clusters. The number of
clusters reduces by one so that the average size of the remaining
clusters (and thus, their stability) marginally increases~\cite{Tang87}.
Their growing size naturally decelerates the dynamics. The algorithm
is exceedingly simple and consists of only \emph{two} choices, yet,
it readily reproduces experimental data~\cite{Robe16}: Particles
always completely fill a lattice, one on each site, but each particle
either (1) is mobile ($h=1$), or (2) it is locked in a cluster of
size $h>1$ with adjacent particles. At the time of quench ($t=0$),
all particles are mobile. When picked for an update at time $t>0$,
(1) a mobile particle with $h=1$ swaps position with a random neighbor
and joins its cluster, conversely, (2) a particle in a cluster of
size $h>1$ breaks up the entire cluster with probability $P(h)$.
While Ref.\ \cite{Becker14} already obtained a hyperbolic event
rate for the cluster model, consistent with Fig.\ \ref{fig:Rate}b-c,
in Fig.\ \ref{fig:ClusterPoisson} we demonstrate its log-Poisson
behavior.

In conclusion, our study shows that existing models of aging based
on renewal processes are inconsistent with the physical evidence of
aging exhibiting `jumps' or `quakes'  describable as a 
log-Poisson process. The implications of this finding
for other models of aging remain less obvious. Amir et al\ \cite{Amir2012,Lahini2017}
stipulate a convolution of relaxation rates $\lambda$ with distribution
$P(\lambda)\sim\lambda^{-1}$ that implies aging of observables $\sim\log(t/t_{w})$,
consistent with our description. Earlier theories\ \cite{Cugliandolo93,Vincent96},
derived from mean-field spin glasses before experiments implicated
the importance of intermittency~\cite{Bissig03,Sibani06a,Yunker09,Kajiya13,Tanaka17},
describe aging merely as a gradual process. Future analysis will reveal
its consistency with the evidence in its system-wide averages, but
it lacks any notion of \emph{localized} spatiotemporal heterogeneity
now considered essential in the understanding of slow relaxation.
In contrast, a description of jamming in terms of a random first-order
transition (RFOT)\ \cite{Lubchenko2004,Lubchenko2017} explains the
mechanics of \emph{individual} cage-breaking events in structural
glasses in great detail, putting some emphasis on the irreversibility
of the event and the structural relaxation it implies. There, the
increasing free-energy barriers we stipulated for reaching lower metastable
basins in the landscape are explained in terms of an impending entropy crisis:
lower-energy basins are ever harder to find. Yet, we are not aware
of any prediction within RFOT about the observed $1/t$-deceleration
in the rate of quake events, or any other hallmark of a log-Poisson
process. In fact, the need for such a microscopic justification for
aging is somewhat antithetical to the rather broad universality found
here for both structural as well as quenched glasses, at least for
an elementary protocol.

\bibliographystyle{apsrev4-1}
\bibliography{/Users/sboettc/Boettcher}

\end{document}